\documentclass[11pt,preprint]{aastex}

\def\dg{{\rm o}}
\def\be{\begin{equation}}
\def\ee{\end{equation}}

\def\Sgr{Sgr~A$^*$~}

\def\gav{\bar{g}}

\def\Mtrue{M_{\rm true}}
\def\rtrue{r_{\rm true}}
\def\vtrue{v_{\rm true}}
\def\R0true{R_{0\rm true}}
\def\Mtrial{M_{\rm trial}}
\def\bv{{\mathbf v}}
\def\br{{\mathbf r}}
\def\bl{{\mathbf l}}
\def\bn{{\mathbf n}}

\newbox\grsign \setbox\grsign=\hbox{$>$} \newdimen\grdimen \grdimen=\ht\grsign
\newbox\simlessbox \newbox\simgreatbox \newbox\simpropbox
\setbox\simgreatbox=\hbox{\raise.5ex\hbox{$>$}\llap
     {\lower.5ex\hbox{$\sim$}}}\ht1=\grdimen\dp1=0pt
\setbox\simlessbox=\hbox{\raise.5ex\hbox{$<$}\llap
     {\lower.5ex\hbox{$\sim$}}}\ht2=\grdimen\dp2=0pt
\setbox\simpropbox=\hbox{\raise.5ex\hbox{$\propto$}\llap
     {\lower.5ex\hbox{$\sim$}}}\ht2=\grdimen\dp2=0pt
\def\simgt{\mathrel{\copy\simgreatbox}}
\def\simlt{\mathrel{\copy\simlessbox}}

\begin{document}

\title{Clockwise stellar disk and the dark mass in the Galactic Center}

\author{Andrei M. Beloborodov\altaffilmark{1,2}, Yuri Levin\altaffilmark{3},
        Frank Eisenhauer\altaffilmark{4}, Reinhard Genzel\altaffilmark{4,5}, 
        Thibaut Paumard\altaffilmark{4}, Stefan Gillessen\altaffilmark{4}, 
        Thomas Ott\altaffilmark{4}} 

\altaffiltext{1}{Physics Department and Columbia Astrophysics Laboratory,
Columbia University, 538  West 120th Street New York, NY 10027}

\altaffiltext{2}{Astro-Space Center of Lebedev Physical 
Institute, Profsojuznaja 84/32, Moscow 117810, Russia} 

\altaffiltext{3}{Leiden Observatory, PO Box 9513, NL-2300 RA, Leiden, 
The Netherlands}

\altaffiltext{4}{Max-Planck Institut f\"ur extraterrestrische Physik (MPE), 
Garching, Germany}

\altaffiltext{5}{Department of Physics, University of California, Berkeley, 
CA 94720}

\begin{abstract}
Two disks of young stars have recently been discovered in the 
Galactic Center. The disks are rotating in the gravitational field 
of the central black hole at radii $r\sim 0.1-0.3$~pc and thus open 
a new opportunity to measure the central mass. We find that the 
observed motion of stars in the clockwise disk implies 
$M=(4.3\pm 0.5)\times 10^6M_{\odot}$ for the fiducial distance to the 
Galactic Center $R_0=8$~kpc and derive the scaling of $M$ with $R_0$.
As a tool for our estimate we use orbital roulette, a recently 
developed method.  The method reconstructs the three-dimensional orbits 
of the disk stars and checks the randomness of their orbital phases.
We also estimate the three-dimensional positions and orbital eccentricities 
of the clockwise-disk stars.
\end{abstract}



\section{INTRODUCTION}

A supermassive black hole is believed to reside in the dynamical center of 
the Galaxy (Eckart \& Genzel 1996; Genzel et al. 1997; Ghez et al. 1998;
Sch\"odel et al. 2002;
Genzel et al.~2000; Ghez et al.~2000; 
Sch\"odel et al.~2003; Ghez et al.~2003). It is associated with the compact 
radio source \Sgr and dominates the gravitational potential at distances 
$r< 1$~pc from the center. 
The central mass has been estimated with a few different methods. 
In particular, the statistical analysis of sky-projected positions and 
velocities of stars in the central 0.5~pc gave estimates that range
above $3\times 10^6M_\odot$ (Genzel et al. 2000; Sch\"odel et al.~2003).
The systematic error of these estimates, however, is difficult to 
quantify as they depend on the assumed three-dimensional statistical model 
of the cluster. 

Independent estimates of the black hole mass were derived
from observations of a few stars (S-stars) that move on very tight orbits 
around \Sgr, at radii $r\sim 0.01$~pc, and have orbital periods as 
short as a few decades. Fractions of S-star orbits have been mapped out over 
the past decade (Sch\"odel et al.~2002; Sch\"odel et al.~2003; 
Ghez et al.~2003; Ghez et al.~2005). By checking which value of the 
central mass provides an acceptable fit to the orbital data, 
Eisenhauer et al. (2005) found $M=(4.1\pm 0.4)\times 10^6M_\odot$, and 
Ghez et al. (2005) found $M=(3.7\pm 0.2)\times 10^6M_\odot$ for 
the fiducial distance to the Galactic Center $R_0=8$~kpc.
Same star orbits were used to derive $R_0=7.62\pm 0.32$~kpc 
(Eisenhauer et al. 2005), in agreement with previous methods 
(e.g. Reid et al. 1993). 

In this paper, we report a new independent estimate of the central 
mass that uses the young massive stars at $r\sim 0.1-0.3$~pc as 
test particles. A remarkable 
property of these stars has been discovered recently: they form a 
disk-like population (Levin \& Beloborodov 2003; Genzel et al. 2003). 
Two stellar disks apparently coexist at these radii. One of them 
rotates clockwise on the sky and the other one --- counter-clockwise.
New observations of Paumard et al. (2006) have increased the 
number of known disk members by a factor of three and show clearly the 
two-disk structure of the young population. The clockwise disk is 
especially well pronounced. It has about 30 members and a small 
thickness $H/r \approx 0.1$. 

In this paper, we investigate the orbital motions of stars in the 
clockwise disk and infer the central gravitating mass. 
As a tool of our mass estimate we use orbital roulette, a recently 
developed statistical method (Beloborodov \& Levin 2004, hereafter BL04). 
This method is generally more precise than the estimators based 
on the virial theorem and provides a statistically well-defined way to 
quantify the uncertainty of the mass estimate.


\section{CLOCKWISE DISK}

The disk-like structure of the young stellar population is not evident
when it is observed in projection: the line-of-sight coordinaties $z$
of the stars are unknown. This structure, however, is evident in the
distribution of the measured 3D velocities. In particular, the velocity 
vectors of stars that move clockwise on the sky cluster near a particular 
plane. This may be only if the orbits of stars are near this plane 
and form a disk-like population. 

The clockwise disk
has a small thickness at angular distances $0.7"<d<5"$ from \Sgr. We use
the stars in this region to define the midplane of the disk. It is defined
as the midplane of the observed velocity vectors, and its normal vector 
is\footnote{$(x,y,z)$ are standard coordinates: they are
chosen so that \Sgr is at the origin, the $x$ and $y$ axes are in the
plane of the sky and directed along increasing right ascension and
declination, respectively, and the $z$ axis is directed along the line
of sight away from us.} 
$\bn_0=(-0.135,-0.857,0.497)$. The deviations of individual 
stellar velocities from the midplane are shown in Figure~1. 
Hereafter in this paper we use the sample of stars between 0.7" and 5" 
with firmly detected clockwise rotation $l_z=xv_y-yv_x>2\Delta l_z$,
where $\Delta l_z$ is the measurement error in $l_z$.

\begin{figure}
\begin{center}
\plotone{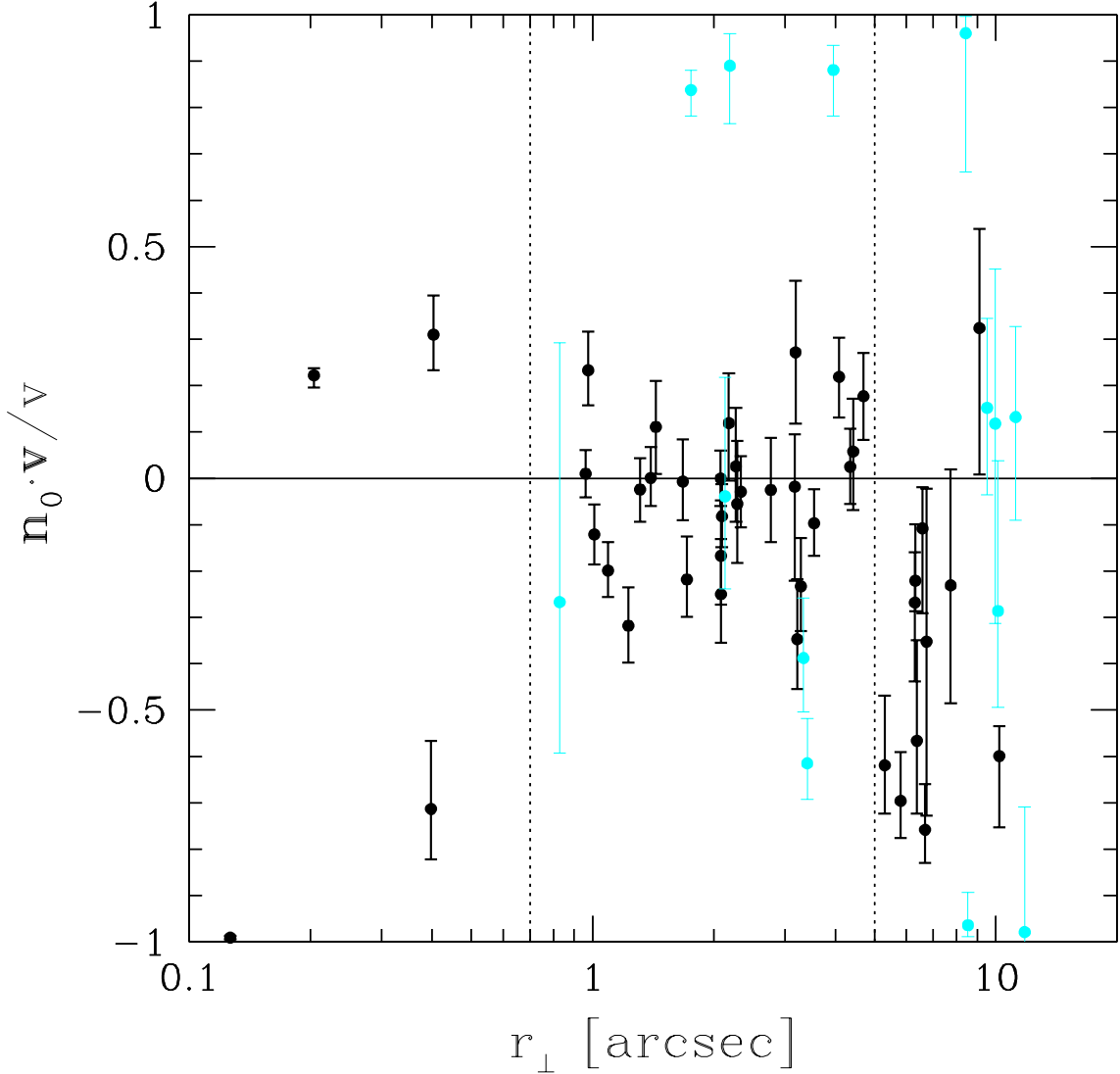}
\end{center}
\caption{
Quantity $\mu={\mathbf n}_0\cdot{\mathbf v}/v$ for all young stars 
observed by Paumard et al. (2006) 
that have apparent clockwise rotation on the sky $l_z>0$, shown
versus projected distance from \Sgr.
For randomly oriented orbits, $\mu$ would be uniformly distributed
between $-1$ and $1$. The actual data show a strong clustering around
$\mu=0$ (midplane). Data shown in black represent the stars with
firmly detected clockwise rotation $l_z>2\Delta l_z$, and 
cyan points --- the stars with $0<l_z<2\Delta l_z$, where $\Delta l_z$ 
is the error in $l_z$.
The sample used in this paper is the 28 stars shown in black in the
region $0.7" < r_\perp < 5"$.
}
\end{figure}

\begin{figure}
\begin{center}
\plotone{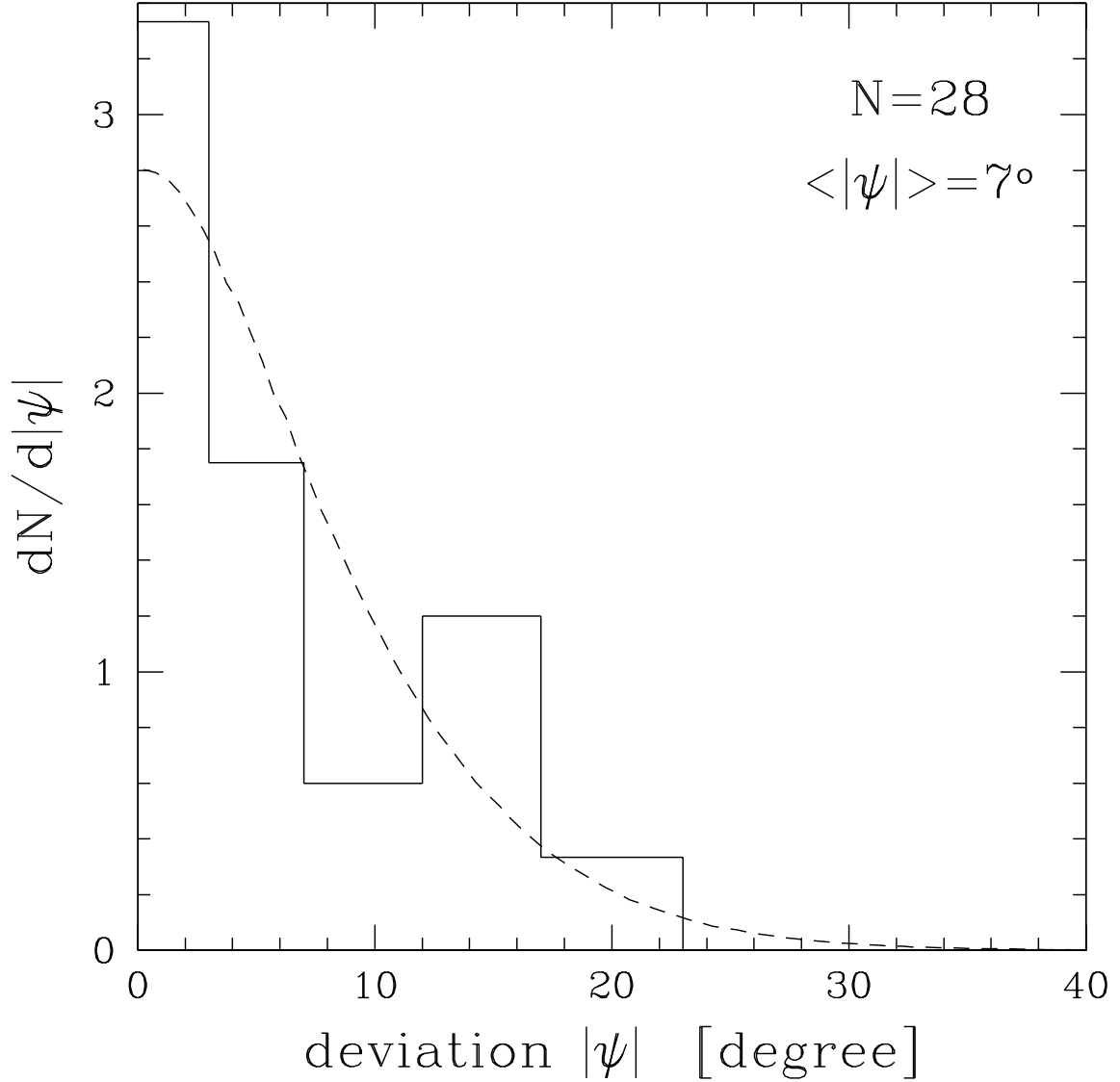}
\end{center}
\caption{
Histogram of the angles $\psi_i$ betwen observed best-fit velocities
$\bv_i$ and the disk midplane with normal ${\mathbf n}_0$. 
Angle $\psi$ is related to $\mu$ shown in Figure~1 by $\sin\psi=\mu$.
The mean value of $|\psi|$ is $\langle |\psi|\rangle=7.1^\dg$ and 
$\langle\psi^2\rangle^{1/2}=9.3^\dg$.
Dashed curve shows the distribution of $|\psi|$ expected from our
model disk population. The dispersion of orbital inclinations $\theta$ 
in the model is $10^\dg$.
}
\end{figure}

\subsection{Thickness of the Disk}

The sample has 28 stars, which is sufficiently large to study the
distribution of velocity deviations from the disk midplane (Fig.~2).
The deviation angles $\psi_i$ are defined by 
$\psi_i=\arcsin(\bn_0\cdot\bv_i/v_i)$.
The observed distribution of $\psi$ may be fitted
by a disk population model. We will use a simple Gaussian model 
of orbital inclinations $\theta$ around the midplane. The probability 
distribution of $\theta$ is then given by 
$P(\theta)=(2/\pi)^{1/2}(\Delta\theta)^{-1}\exp[-(\theta/\Delta\theta)^2/2]$. 

The Gaussian model of the disk has one parameter --- the dispersion 
$\Delta\theta$ --- and predicts a certain probability distribution for $\psi$. 
We calculate this distribution using Monte-Carlo technique, including a
simulation of observational errors. We randomly draw $N=28$ orbits 
from the model population and choose a randomly directed vector of length
$v_i$ in each orbital plane $i=1,...,N$.
This gives 28 simulated velocities $\bv_i$. Then observational 
errors $\Delta \bv_i$ are added to each $\bv_i$ (the real $v_i$ and
$\Delta\bv_i$ from Paumard et al. 2006 are used). Finally, deviations 
$\psi_i$ of velocities
$\bv_i$ from the disk midplane are calculated. This simulation is repeated 
$10^5$ times, which gives the accurate distribution of $\psi$ 
predicted by the model as well as the average values $\langle|\psi|\rangle$ 
and $\langle\psi^2\rangle^{1/2}$ as functions of $\Delta\theta$.

The observed $\langle|\psi|\rangle\,\approx 7^\dg$ and 
$\langle\psi^2\rangle^{1/2}\approx 9^\dg$ 
are reproduced by the Monte-Carlo model with $\Delta\theta\approx 10^\dg$. 
The model is consistent with the observed distribution of $\psi$ for
$\Delta\theta=10\pm 3^\dg$. Hereafter we will use the best-fit model with 
$\Delta\theta=10^\dg$.

The geometrical thickness of the disk may be described by 
the elevation of stars above/below the midplane, 
\be
          \frac{H}{r} = |\sin\phi\sin\theta|,
\ee
where $\phi$ is the orbital phase of the star, and $\theta$ is the 
inclination of its orbit.
Assuming circular orbits for simplicity, one finds after averaging
over the orbital phase,
\be
        \frac{H}{r} = \left(\frac{2}{\pi}\right)^{1/2}\langle\sin\theta\rangle.
\ee
This gives for the Gaussian model (approximating 
$\sin\langle\theta\rangle\,\approx\,\langle\sin\theta\rangle$ for small 
$\Delta\theta=0.17$),
\be
\label{eq:H}
  \frac{H}{r}=\left(\frac{2}{\pi}\right)^{3/2}\sin\Delta\theta
             =0.09\pm 0.03.
\ee

\subsection{Circular Motion?}

We now check whether the data is consistent with 
circular motion of the disk stars.
The assumption of circular motion implies that radius-vector
$\br_i$ of the $i$-th star satisfies
\be
   \br_i^{\rm cir}\cdot\bv_i=0,  \qquad i=1,...,N. 
\ee
This equation, together with the known position ($x_i,y_i$) on the sky, 
determines $z_i$ --- the line-of-sight coordinate of the star.
The easiest way to test the circular-motion hypothesis
is to look at the orientations of orbital planes implied by 
this hypothesis. The inferred $\br_i^{\rm cir}$ allow us to calculate 
the angular momentum vector for each star
\be
  \bl_i^{\rm cir}=\br_i^{\rm cir}\times\bv_i, \qquad i=1,...,N,
\ee
and then find its deviation $\theta^{\rm cir}_i$ from the disk axis $\bn_0$.
The dispersion of $\theta^{\rm cir}_i$
would be expected to be consistent 
with disperion $\Delta\theta\approx 10^\dg$ of orbital planes inferred 
directly from $\bv_i$.
The actual distribution of $\theta^{\rm cir}_i$ is shown in Figure~3.
One can see that the circular hypothesis implies orbital inclinations that 
are inconsistent with the found thickness of the disk. 
A significant fraction of the disk stars must have non-negligible 
eccentricities.

\begin{figure}
\begin{center}
\plotone{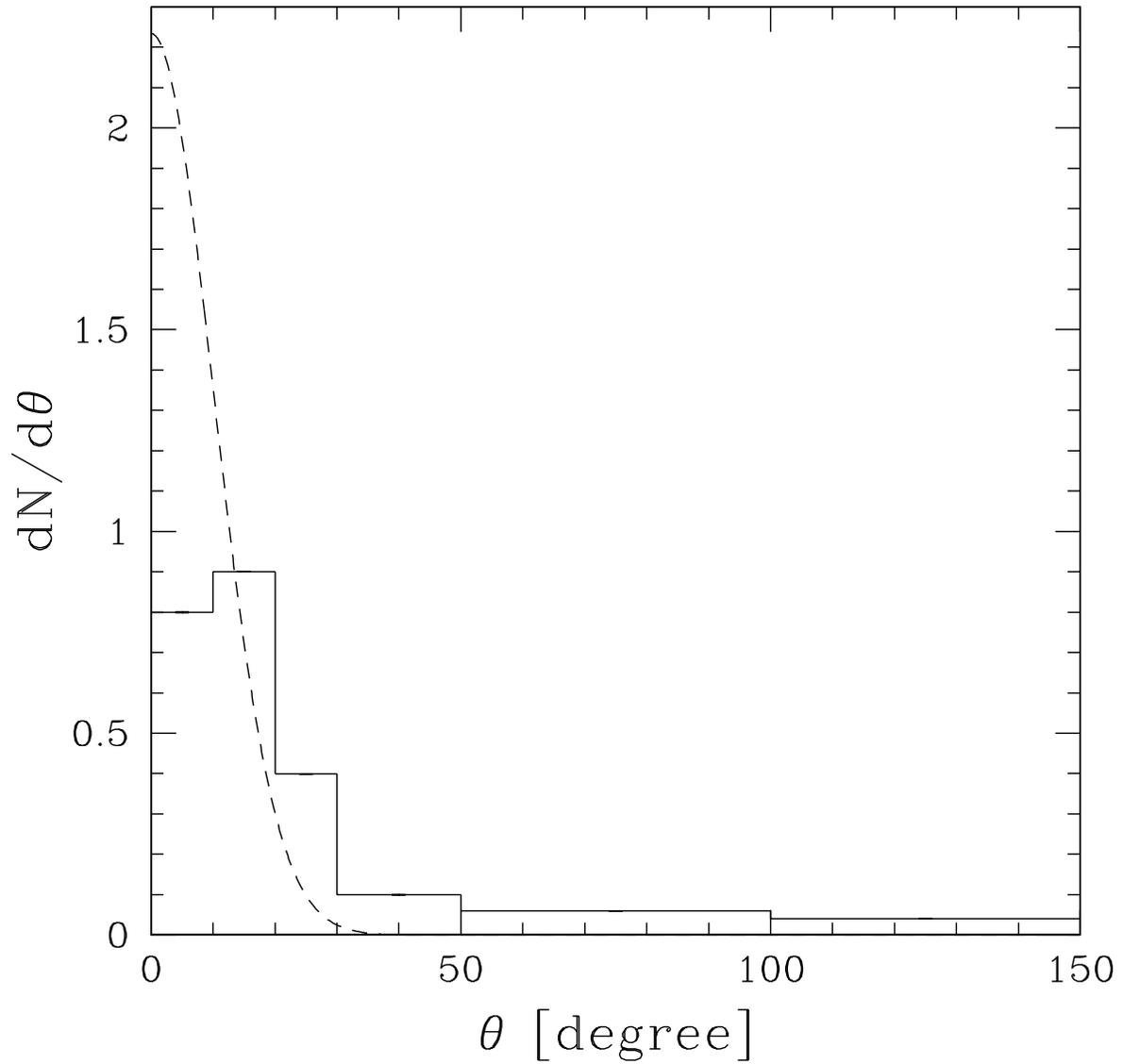}
\end{center}
\caption{
Histogram of orbital inclinations $\theta$ that would be 
inferred from the circular hypothesis. Dashed curve shows the 
$10^\dg$ Gaussian distribution of $\theta$ that fits the 3D velocity 
data (see Fig.~2). The circular hypothesis 
implies large orbital inclinations $\theta > 30^\dg$ for at least 1/4 
of stars, which is inconsistent with the observed narrow distribution 
of velocity angles $\psi_i$.
}
\end{figure}

A circular orbit would provide a straightforward estimate of the central 
mass $M$.
Indeed, circular motion implies a simple relation between $M$, $v$, and 
$r$: $GM=rv^2$. 
An estimate of $M$ could be done by applying this relation
to stars with small $\theta^{\rm cir}$, which are possibly on circular orbits.
However, it is difficult to quantify the accuracy of such analysis.
A more accurate estimate of $M$ is made in \S~4, without making any prior 
assumptions on eccentricities.


\subsection{Three-Dimensional Positions}  

Measurements of stellar positions relative to \Sgr are limited to 
2 dimensions --- the plane of the sky. However, the stars that belong to 
the clockwise disk have an extra constraint on their 3D 
positions $\br_i$: they must be close to the midplane of the stellar disk. 
This constraint may be written as
\be
  ({\mathbf n}_0\pm{\Delta\mathbf n})\cdot{\mathbf r}_i=0, \qquad  i=1,..,N,
\ee
where ${\mathbf n}_0$ is the disk axis and $\Delta {\mathbf n}$ represents 
the dispersion of orbital planes around the disk midplane.

The line-of-sight coordinate of the $i$-th star is then given by
\begin{equation}
  z_i=-\frac{x_i n_{x,i}+y_i n_{y,i}}{n_{z,i}}, \qquad 
     {\mathbf n}_i={\mathbf n}_0\pm{\Delta\mathbf n}.
\label{zi}
\end{equation}
The orbital plane ${\mathbf n}_i$ may differ from the 
best-fit disk plane by $\sim 10^\dg$, which induces an error in the 
inferred $z$-coordinate. To estimate the error we rock ${\mathbf n}_i$ 
with a Gaussian distribution around $\bn_0$ with a mean deviation of 
$10^\dg$. 

The resulting $z_i$ and their uncertainties are found using
the following Monte-Carlo simulation.
We draw randomly ${\mathbf n}_i$ and $v_{x,i}$, $v_{y,i}$ from their
measured Gassian distributions. Then we determine $v_{z,i}$ from the
condition ${\mathbf v}_i\cdot{\mathbf n}_i=0$.
The obtained realization of ${\mathbf n}_i$ and ${\mathbf v}_i$ 
is given a weight that corresponds to the level of consistency of $v_{z,i}$ 
with the measured Gaussian $v_{z,i}=v_{z,i}\pm\Delta v_{z,i}$.
(In the simulation, we get this weight by rejecting the realization 
if its $v_{z,i}$ is farther in the Gaussian tail than a randomly chosen 
$v_{z,i}^\prime$ from the Gaussian distribution.)
In this way, a realization ${\mathbf n}_i$, ${\mathbf v}_i$ is given
the weight that takes into account both 
the thickness of the disk and the error of the 3D velocity.
We find the mean $z_i$ and its dispersion $\Delta z_i$ using about $10^6$ 
realizations.

 \begin{figure}
 \begin{center}
 \plotone{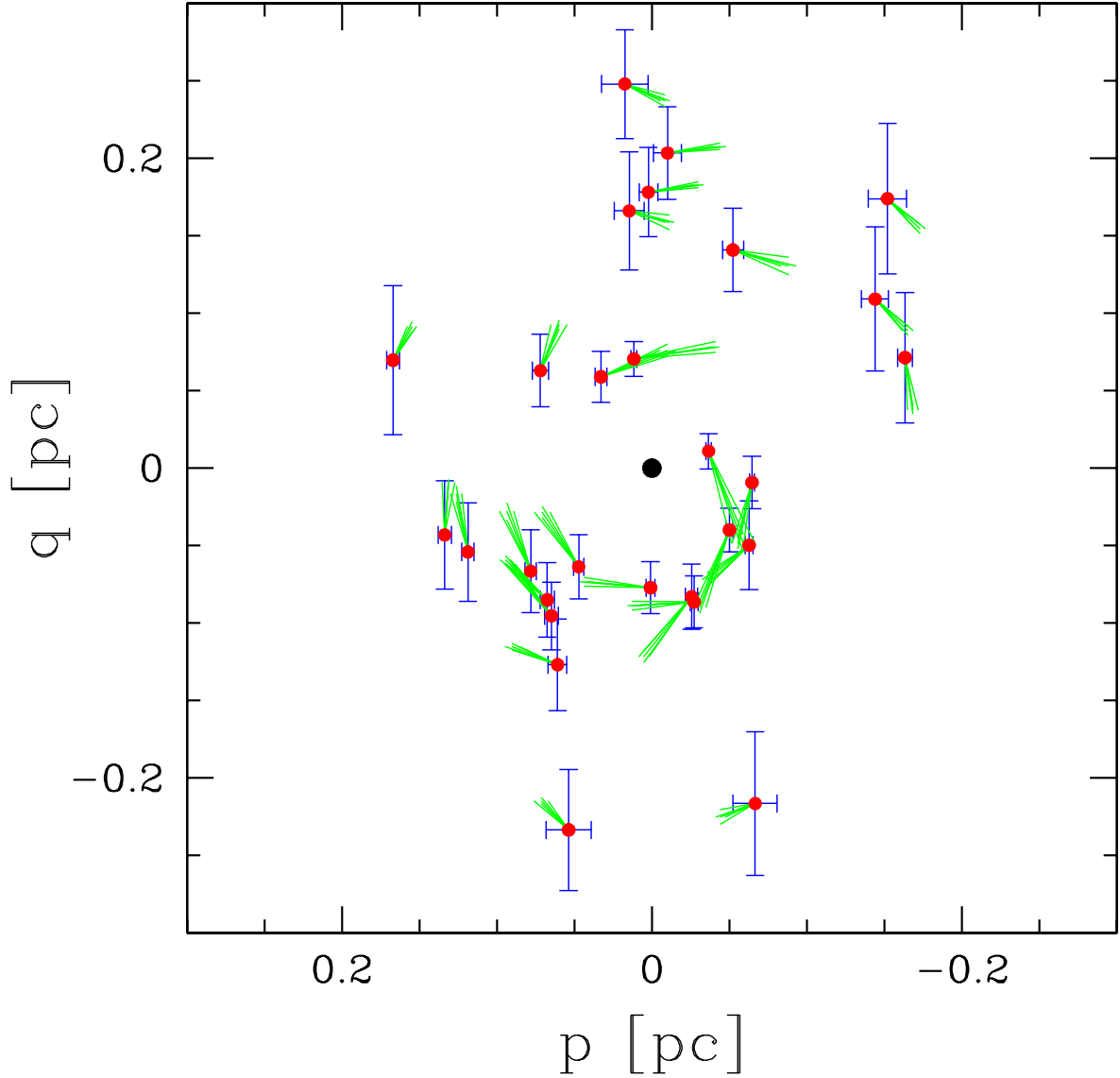}
 \end{center}
 \caption{
Positions and velocities of the 28 stars in the disk plane $(p,q)$.
\Sgr is located at the origin (black circle).
The $p$-axis is chosen in the intersection of the disk plane with the
$(x,z)$ plane and the $q$-axis is perpendicular to the $p$-axis. 
The axis directions (unit vectors) ${\mathbf e}_p$ and ${\mathbf e}_q$ 
are chosen so that ${\mathbf e}_p\cdot {\mathbf e}_x>0$ and
${\mathbf e}_q\cdot{\mathbf e}_y>0$.
The velocity of each star is shown by 5 vectors: the best-fit vector
${\mathbf v}_i$, ${\mathbf v}_i\pm \Delta v_p{\mathbf e}_p$, and
${\mathbf v}_i\pm \Delta v_q{\mathbf e}_q$.
We show the stars in a single plane, as if they had a common orbital 
plane. Their real orbits may differ from the disk midplane by $\sim 10^\dg$,
and the corresponding uncertainties in positions are indicated in the 
figure. The distance to the Galactic Center $R_0=8$~kpc is assumed.
 }
 \end{figure}

The deduced 3D positions together with the measured 3D velocities give 
6D information for each star. Then the velocities $v_r$ and $v_\phi$ may
be deduced, where $r$ and $\phi$ are polar coordinates in the 
orbital plane of the star.
The measurement errors of $\bv_i$ and the uncertainty in the individual 
orbital planes imply errors $\Delta v_r$ and $\Delta v_\phi$.
We estimate the uncertainties 
assuming the Gaussian dispersion in ${\mathbf n}_i$ with $\Delta\theta=10^\dg$ 
and Gaussian velocity errors ${\Delta \mathbf v}_i$. 
The results are summarized in Table~1 and illustrated in Figure~4.


\section{MASS ESTIMATION USING ORBITAL ROULETTE}

The orbital periods of stars at $r\simgt 0.1$~pc are thousands of years, and 
it is presently impossible to trace any significant fraction of the orbits. 
Therefore, a statistical method must be applied using the observed 
{\it instantaneous} positions $\br_i$ and velocities $\bv_i$ of the stars. 
Such a method must take into account the possible errors in $\br_i$ and
$\bv_i$.

A traditional mass estimate would be based on the virial theorem,
\be
\label{eq:vir}
   M=\frac{\sum v_i^2}{G\sum 1/r_i}.
\ee
The virial relation gives $M=3.8\times 10^6M_\sun$.
One faces difficulties, however, when trying to define the error 
of this estimate (see BL04 for discussion).

A major assumption in the derivation of the virial relation~(\ref{eq:vir})
is that the stars have random orbital phases. BL04 have shown that the 
random-phase (``roulette'') principle can be applied directly to data,
without invoking the virial theorem, and this gives a more accurate and 
statistically complete estimator.
It judges any trial gravitational potential $\Phi(r)$ 
by reconstructing the orbits in this potential and checking the inferred 
orbital time phases. The true potential must give a ``fair roulette'' 
--- a random (flat) distribution of the phases. 

The 3D positions of the disk stars have uncertainties because their 
precise orbital planes are unknown. Besides, the 3D velocities are 
measured with errors. We take into account the uncertainties using Monte-Carlo
technique: many random realizations of the data set $(\br_i,\bv_i)$ are 
generated and studied. The possible $(\br_i,\bv_i)$ are generated assuming 
the $10^\dg$ Gaussian distribution of the orbital planes and Gaussian errors 
in $\bv_i$ with dispersions $\Delta\bv_i$ given in Paumard et al. (2006).

Each realization of the data set $(\br_i,\bv_i)$ is analyzed by the 
roulette estimator as described in BL04. The analysis aims
to accept or rule out a trial potential $\Phi(r)=-G\Mtrial/r$ 
at a chosen confidence level. Given $\Mtrial$ and a data set 
$(\br_i,\bv_i)$ one can reconstruct all $N=28$ orbits 
and find the time orbital phases $g_i$ of the stars. 
The time phase $g$ is defined so that $g=0$ at the pericenter 
of the orbit and $g=1$ at the apocenter. To judge $\Mtrial$ we
check the consistency of $g_i$ with the expected flat 
distribution between 0 and 1 (and do that for many realizations of the 
data sets to take into account the uncertainties in the data).

The roulette algorithm is described in detail in BL04. 
Here we use its simplest version based on the analysis of the mean phase
$\gav=N^{-1}\sum g_i$. The true $M$ must satisfy the condition 
\be
\label{eq:roul}
\bar{g}(M)=0.5\pm(12N)^{-1/2}.
\ee
If $\Mtrial$ is too small then $\bar{g}(\Mtrial)\rightarrow 0$, i.e. all 
stars will be found near the pericenter, while too large $\Mtrial$ 
will give $\bar{g}(\Mtrial)\rightarrow 1$, i.e. all stars will sit near 
apocenter. One can accept only such $\Mtrial$ that give $\gav$ consistent 
with the fair roulette (eq.~\ref{eq:roul}).

In our Monte-Carlo simulation, $\gav(\Mtrial)$ that is found for a 
realization of the data set $(\br_i,\bv_i)$ is compared with $\gav^\prime$ 
found for a realization of a fair roulette $g_i^\prime$. 
The statistics of this comparison is accumulated for $10^6$ random 
realizations. Thus, the probability $P_-$ of 
$\gav(\Mtrial)<\gav^\prime$ is accurately found. 

The probability $P_-(\Mtrial)$ is monotonically increasing with $\Mtrial$
since larger masses imply larger orbital phases (closer to the apocenter).
Trial masses that give small $P_-$ are ruled out with significance 
$P_-$ (BL04). If $P_->1/2$, we look at $P_+=1-P_-$. Trial masses 
that give small $P_+<1/2$ are ruled out with significance $P_+$. 

The found functions $P_\pm(\Mtrial)$ are shown in Figure~5.
The confidence intervals for $M$ are between the two curves.
For example, the 80\% confidence interval ($P_-=P_+=0.1$)
is $3.85\times 10^6M_\odot < M < 4.93 \times 10^6M_\odot$.
The confidence intervals approximately correspond to a Gaussian distribution
with standard deviation $(0.4-0.5)\times 10^6M_\odot$, and we can summarize
the result as $M=(4.3^{+0.5}_{-0.4})\times 10^6M_\odot$. 

We have also checked the distribution of $g_i$ for the found $M$. 
It is consistent with the expected flat distribution; this is verified 
using the Anderson-Darling measure (see BL04). We conclude that the 
observed stellar motions are consistent with Kepler potential $\Phi=-GM/r$ 
with $M=(4.3^{+0.5}_{-0.4})\times 10^6M_\odot$.

There are two sources of systematic error of our result. First, we had
to chose a disk-population model in the calculations. We specified 
the midplane (normal $\bn_0$)
and the Gaussian distribution of orbital 
inclinations around the midplane with dispersion $\Delta\theta$.
The parameters $\bn_0$ and $\Delta\theta$ have been derived by fitting 
the 3D velocity data and have finite errors.
The obtained $M$ 
is most sensitive to $n_{z0}$ which determines the disk inclination to 
the line of sight. Our best-fit disk has $n_{z0}\approx 0.5$;
larger $n_z$ would imply smaller sizes of the orbits and
hence smaller $M$. $n_{z0}$ as
large as $0.6$ may still be consistent with the velocity data (Paumard 
et al. 2006), and it would change our estimate to 
$M=(4.0^{+0.5}_{-0.3})\times 10^6M_\odot$. The parameter $\Delta\theta$ has 
the error of $3^\dg-4^\dg$. The choice of
$\Delta\theta=13^\dg$ instead of $10^\dg$ would broaden the confidence
intervals by a factor $\sim 1.2$.

The second possible source of error is the distance to the Galactic
Center $R_0$ which we assumed to be 8~kpc. The mass estimate depends on 
$R_0$ because the proper velocities $\bv_\perp=(v_x,v_y)$ and 
positions $(x,y)$ scale linearly with $R_0$ ($v_z$ is not affected as
its measurement is based on the Doppler effect). The midplane of the disk 
varies little and may be assumed constant to a first approximation.
Then the deduced radial positions $r$ of the stars scale linearly with $R_0$.
The dependence of $M$ on $R_0$ may be derived using a toy problem with 
$N\gg 1$ stars moving in the disk midplane on a circular orbit of 
radius $\rtrue$. The true and deduced masses are then given by 
$$
 G\Mtrue = \rtrue \left(v_{\perp,\rm true}^2 + v_z^2 \right),
$$
$$
 GM = r \left(v_{\perp}^2 + v_z^2 \right)
    =\rtrue \left(\frac{R_0}{\R0true}\right) 
   \left[v_{\perp,\rm true}^2\left(\frac{R_0}{\R0true}\right)^2 + v_z^2\right].
$$
After averaging over $N$ stars and taking into account that 
$\langle v_{\perp,\rm true}/\vtrue\rangle = (1 + n_z^2)/2$ and
$\langle v_z/\vtrue\rangle =(1 - n_z^2)/2$, we find
\be
\label{eq:scale}
  M(\R0true)=M(8{\rm ~kpc})\left[\frac{1}{2}(1 + n_z^2)
                \left(\frac{\R0true}{8\rm ~kpc}\right)^{-3}
         + \frac{1}{2}(1 - n_z^2)\left(\frac{\R0true}{8\rm ~kpc}\right)^{-1}
                \right]^{-1}.
\ee
We have empirically checked this dependence by repeating the 
roulette simulation for various $R_0$ (as small as 7~kpc).
The obtained $P_\pm(M)$ are well described by a simple shift of 
$M$ according to equation~(\ref{eq:scale}). In particular, 
$R_0=7.6$~kpc (Eisenhauer et al. 2005) would give
$M=(3.8^{+0.4}_{-0.3})\times 10^6M_\odot$.

\begin{figure}
\begin{center}
\plotone{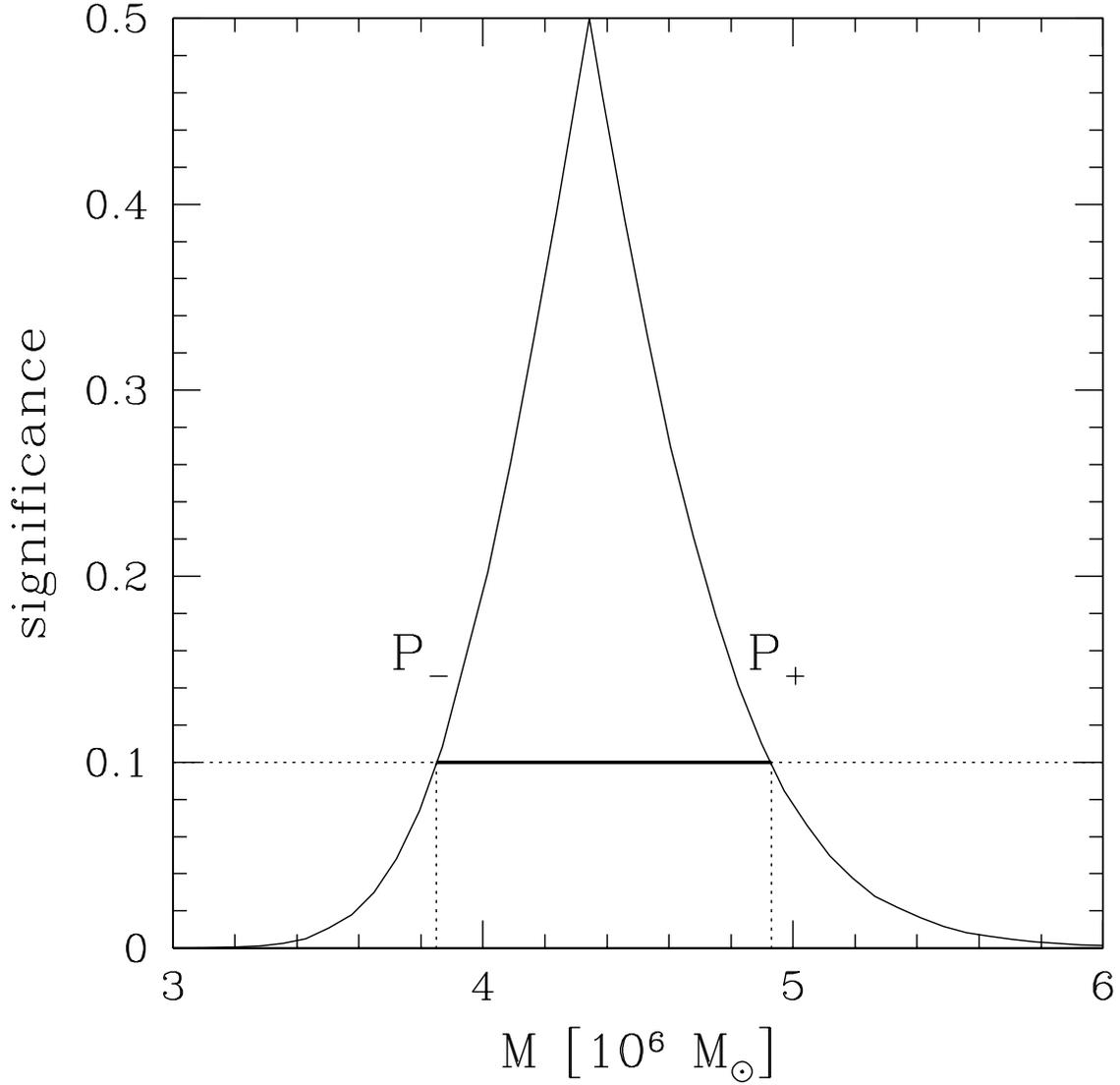}
\end{center}
\caption{ Confidence intervals for \Sgr mass $M$ found with the roulette 
method. For example the 80\% confidence interval ($P_+=P_-=0.1$) is shown 
by the horizontal solid line ($P_+=P_-=0.1$). 
The distance to the Galactic Center $R_0=8$~kpc is assumed. 
The scaling of the estimated mass with $R_0$ is given by eq.~(\ref{eq:scale}).
}
\end{figure}

Table~1 shows the eccentricities of reconstructed star orbits for
$R_0=8$~kpc and the best-fit central mass $M=4.3\times 10^6M_\odot$. 
The errors $\Delta e_i$ ($i=1,...,28$)
are caused by uncertainties $\Delta {\mathbf v}_i$ and $\Delta{\bf n}_i$. 
The errors are found using the Monte-Carlo simulation that is similar to 
the estimation of $\Delta z_i$ (see \S~2.3). This simulation gives
the probability distribution $f(e_i)$ for each $e_i$. A non-zero eccentricity 
is detected if $\bar{e}_i>2\Delta e_i$ and accurately measured if 
$\bar{e}_i>3\Delta e_i$, where $\bar{e}_i$ is the mean of $f(e_i)$ and 
$\Delta e_i$ is its dispersion. If distribution $f(e_i)$ is broad (large 
errors), its mean represents the errors rather than the true $e_i$.
Therefore, $\bar{e}_i$ is an accurate estimate of eccentricity only if the 
distribution $f(e_i)$ is sufficiently narrow, $\Delta e_i\simlt 0.3\bar{e}_i$. 
Table~1 shows $\bar{e}_i\pm\Delta e_i$ for stars with 
$\Delta e_i< 0.5\bar{e}_i$.  
For stars with $\Delta e_i>0.5\bar{e}_i$ we give 
$\bar{e}_i+\Delta e_i$ as an upper limit.
The obtained values are consistent with the results of Paumard et al. (2006)
where eccentricities are estimated with another method.


\section{DISCUSSION}

The reconstructed orbits of stars in the clockwise disk depend on 
the assumed central mass $M$. In particular, the star orbital phases
depend on $M$. We have used this dependence to estimate $M$: the true 
$M$ must give time phases consistent with a uniform distribution.
This is the first practical application of orbital roulette, a new method 
of estimating the gravitational potential from instantaneous motions of 
test bodies (BL04). 

The obtained estimate $M=(4.3^{+0.5}_{-0.4})\times 10^6M_{\odot}$ is consistent 
with the recent estimates based on the maps of orbits of S stars in the 
very vicinity of the center ($r<0.01$~pc). The independent estimate 
presented here uses different stars at $r=0.1-0.3$~pc and is complementary 
to the small-scale estimates. It measures the total mass within $r\sim 0.1$~pc
which may include stars and dark matter. 

The approximate equality of the masses within the central 0.1~pc and 0.01~pc
implies an upper bound on the mass between 0.01 and 0.1~pc:
$\Delta M \simlt 0.5\times 10^6M_\odot$. This constrains
a possible cusp of the stellar population around the cenral black hole
as well as the dark-matter content of the central region. 
Further observations may provide a better constraint or resolve $\Delta M$.

The data shows an inner cutoff of the disk population at a radius 
$r\sim 0.05$~pc (see Fig.~1 and Paumard et al. 2006).
The observed peak of the population at $r\sim 0.1$~pc supports the hypothesis 
that the young stars were born in a gravitationally unstable accretion disk 
(LB03). 

The found geometrical thickness of the clockwise stellar population 
$H/r=0.09\pm 0.03$ (eq.~\ref{eq:H}) does not imply that it was born with 
this $H/r$. The parent accretion disk was likely thinner, and its stellar 
remnant have thickened with time. The age of the stellar population,
$t_{age}=(6\pm 2)$ million years, should be compared with the timescale
of diffusion of orbital parameters. The latter may be written as 
\be
  t_{diff}\sim 0.1 \frac{v^3}{G^2m_*\rho_*\ln(N_*/2)},
\ee 
where $m_*$ is a characteristic mass of the objects that dominate 
fluctuations of the gravitational potential, $\rho_*$ is their mass
density averaged over the considered region $r\sim 0.3$~pc, and 
$N_*\sim (4\pi/3)r^3\rho_*/m_*$ is the total number of these objects. 
In a uniform stellar core of total mass $\sim 3\times 10^5M_\odot$, 
the fluctuations would be dominated by low-mass stars, whose total number 
is $N_*\sim 10^6$. 
Substituting the observed characteristic velocity, 
$v\sim 300$~km/s, one would find $t_{diff}\sim 10^{10}$~yr$\gg t_{age}$.

However, the Galactic Center is not uniform.
The two stellar disks and stellar clusters are observed, which 
significantly perturb the gravitational potential. These perturbations
should strongly reduce $t_{diff}$. For example, if 1/30 of the core mass
resides in $N_*\sim 10$ objects with $m_*\sim 10^3M_\odot$ then 
$t_{diff}\sim 10^8$~yr. The corresponding
acquired thickness of the stellar disk is 
$H/r\sim t_{age}/t_{diff}\sim 0.1$ in agreement with the observed value.
A similar estimate for $t_{diff}$ may be derived from the interaction 
between the two young disks (Nayakshin et al. 2005).

The eccentricities of disk stars must also diffuse with time.
If the stars were born on circular orbits then $e\sim 0.1$ 
are expected to develop as the disk thickens to $H/r\sim 0.1$.
Table~1 shows that a significant fraction of the young stars in the 
clockwise disk have relatively large eccentricities.
The data are at best marginally consistent with the hypothesis that 
the stars used to have circular orbits 6 million years ago, and
we leave a detailed analysis to a future work. Note that the stars may 
have been born in a gaseous disk that was not completely circularized:
the circularization time was longer than the timescale for star formation. 
Then the initial eccentricities $e$ of the stars are non-zero. 
Besides, the interaction of stars with the disk might have influenced $e$
before the disk was gone.

An alternative scenario assumes that the stars were originally bound to 
an inspiralling intermediate-mass black hole (IMBH; 
Hansen \& Milosavljevic 2003), so that the observed disk 
plane is associated with the orbital plane of the IMBH. 
In this scenario, the inspiralling young cluster was 
stripped by the tidal field of \Sgr and left a disk of stars with a small 
dispersion of orbital planes. Calculations of Levin, Wu, \& Thommes (2005)
show that the stripped stars can generally have eccentric orbits.
However, the inspiralling cluster scenario appears to be in conflict with the 
observed stellar distribution in the clockwise disk (see also Paumard et al. 
2006 for discussion).

\acknowledgements
AMB was supported by Alfred P. Sloan Fellowship.


\begin{table}
\begin{center}
\caption[ ]{Positions and velocities of the clockwise disk stars in their orbital planes}
\end{center}
 {\scriptsize
\begin{tabular}{lrrrrccccc} \hline\hline
&&&&&&&&& \\ [-3.5ex]
 Name                 & $\psi\;\;\;\;$ & $x\;\;\;$ & $y\;\;$ & $z\;\;\;\;\;\;$     &      $r$     &  $\phi$         &    $v_r$     &  $v_\phi$  &  $e$ \\ [0.8ex]
\hline
&&&&&&&& \\ [-3ex]
E15: [GKM98] S1-3     &$  0.5\pm 2.9$&$ 0.52$&$ 1.03$&$ 1.88\pm 0.40$&$ 2.22\pm 0.34$&$  80\pm  3$&$ -12\pm  42$&$ 534\pm 22$&$<0.33$\\[0.8ex] 
E16: [GEO97] W5       &$ 13.7\pm 4.7$&$-1.13$&$ 0.31$&$ 0.27\pm 0.28$&$ 1.23\pm 0.09$&$ 164\pm 16$&$  51\pm 187$&$ 608\pm 63$&$<0.47$\\[0.8ex]
E17                   &$ -6.9\pm 3.8$&$-0.05$&$-1.21$&$-2.04\pm 0.57$&$ 2.40\pm 0.52$&$ 271\pm  2$&$ -33\pm  34$&$ 429\pm 29$&$0.28\pm0.13$\\[0.8ex]
E18: [GEO97] W11      &$-11.4\pm 3.4$&$-0.79$&$-1.04$&$-2.35\pm 0.67$&$ 2.71\pm 0.59$&$ 251\pm  7$&$ 274\pm  68$&$ 408\pm 45$&$0.62\pm 0.14$\\[0.8ex]
E20: GCIRS 16C        &$-18.5\pm 4.9$&$ 1.36$&$ 0.58$&$ 1.47\pm 0.56$&$ 2.11\pm 0.42$&$  59\pm  9$&$ -70\pm  83$&$ 454\pm 48$&$0.34\pm 0.14$\\[0.8ex]
E21: [GEO97] W13      &$ -1.4\pm 4.0$&$-1.02$&$-1.20$&$-2.29\pm 0.54$&$ 2.81\pm 0.50$&$ 252\pm  4$&$ -99\pm  38$&$ 389\pm 29$&$0.35\pm 0.11$\\[0.8ex]
E22: [GEO97] W10      &$  0.2\pm 3.7$&$-1.63$&$-0.37$&$-1.08\pm 0.40$&$ 2.02\pm 0.22$&$ 217\pm 11$&$ 148\pm 106$&$ 505\pm 45$&$0.35\pm 0.15$\\[0.8ex]
E23: GCIRS 16SW       &$  6.4\pm 5.8$&$ 1.26$&$-1.18$&$-1.75\pm 0.61$&$ 2.50\pm 0.48$&$ 308\pm 12$&$-131\pm  85$&$ 398\pm 49$&$0.42\pm 0.16$\\[0.8ex]
E24: [GEO97] W7       &$ -0.2\pm 5.0$&$-2.00$&$ 0.17$&$-0.25\pm 0.46$&$ 2.08\pm 0.13$&$ 188\pm 14$&$ -36\pm  99$&$ 390\pm 43$&$0.46\pm 0.12$\\[0.8ex]
E25: [GEO97] W14      &$-12.3\pm 5.1$&$-1.97$&$-0.60$&$-1.38\pm 0.79$&$ 2.58\pm 0.57$&$ 216\pm 15$&$ -86\pm  94$&$ 329\pm 49$&$0.54\pm 0.15$\\[0.8ex]
E27: GCIRS 16CC       &$ -9.3\pm 6.5$&$ 2.41$&$ 0.65$&$ 1.62\pm 0.65$&$ 3.03\pm 0.39$&$  40\pm 12$&$  95\pm  79$&$ 300\pm 43$&$0.54\pm 0.15$\\[0.8ex]
E28: GCIRS 16SSE2     &$  0.0\pm 3.4$&$ 1.74$&$-1.79$&$-2.55\pm 0.65$&$ 3.61\pm 0.48$&$ 305\pm  9$&$-119\pm  67$&$ 404\pm 31$&$0.37\pm 0.14$\\[0.8ex]
E29                   &$-14.2\pm 6.6$&$ 1.19$&$ 2.20$&$ 3.85\pm 0.68$&$ 5.25\pm 1.36$&$  85\pm  4$&$ -97\pm  49$&$ 252\pm 30$&$0.51\pm 0.13$\\[0.8ex]
E30: GCIRS 16SSE1     &$ -4.7\pm 3.9$&$ 1.91$&$-1.63$&$-2.22\pm 0.69$&$ 3.41\pm 0.52$&$ 310\pm 11$&$ -27\pm  69$&$ 370\pm 28$&$0.27\pm 0.13$\\[0.8ex]
E32: MPE+1.6-6.8      &$  6.4\pm 6.7$&$ 2.22$&$-1.38$&$-1.82\pm 0.76$&$ 3.25\pm 0.47$&$ 321\pm 11$&$-121\pm 101$&$ 387\pm 56$&$0.42\pm 0.18$\\[0.8ex]
E34:MPE+1.0-7.4       &$  1.7\pm 7.1$&$ 1.52$&$-2.26$&$-3.24\pm 0.73$&$ 4.40\pm 0.81$&$ 297\pm  7$&$  27\pm  47$&$ 313\pm 43$&$0.33\pm 0.16$\\[0.8ex]
E35: GCIRS 29NE1      &$ -2.9\pm 7.6$&$-1.19$&$ 2.47$&$ 3.59\pm 0.72$&$ 4.68\pm 0.75$&$ 111\pm  6$&$  29\pm  68$&$ 372\pm 48$&$<0.59$\\[0.8ex]
E36                   &$ -1.7\pm 4.4$&$ 0.54$&$ 2.75$&$ 4.23\pm 0.68$&$ 5.54\pm 0.92$&$  89\pm  2$&$  45\pm  23$&$ 326\pm 29$&$<0.38$\\[0.8ex]
E38                   &$ -1.4\pm 6.5$&$ 0.23$&$ 3.31$&$ 4.16\pm 0.64$&$ 6.35\pm 0.98$&$  93\pm  3$&$  60\pm  25$&$ 336\pm 40$&$<0.68$\\[0.8ex]
E40: GCIRS 16SE2      &$ -3.7\pm 9.4$&$ 3.53$&$-1.43$&$-1.44\pm 0.85$&$ 4.15\pm 0.33$&$ 335\pm 11$&$ -82\pm  95$&$ 365\pm 53$&$0.40\pm 0.19$\\[0.8ex]
E41: GCIRS 33E        &$ 16.0\pm 9.4$&$ 0.78$&$-3.74$&$-4.29\pm 0.60$&$ 7.85\pm 1.65$&$ 283\pm  6$&$-143\pm  32$&$ 214\pm 47$&$0.61\pm 0.14$\\[0.8ex]
E43                   &$-19.8\pm 7.2$&$-1.92$&$-3.35$&$-4.07\pm 0.71$&$ 7.45\pm 1.82$&$ 253\pm  7$&$  16\pm  55$&$ 238\pm 32$&$0.38\pm 0.17$\\[0.8ex]
E44                   &$-13.3\pm 5.9$&$ 1.74$&$ 3.54$&$ 4.36\pm 0.51$&$ 8.14\pm 1.58$&$  86\pm  4$&$-124\pm  40$&$ 252\pm 30$&$0.53\pm 0.16$\\[0.8ex]
E50: GCIRS 16SE3      &$ -5.5\pm 4.1$&$ 4.02$&$-1.39$&$-1.12\pm 0.91$&$ 4.49\pm 0.26$&$ 340\pm 10$&$-121\pm  83$&$ 308\pm 37$&$0.41\pm 0.18$\\[0.8ex]
E54                   &$ 12.6\pm 5.0$&$-4.40$&$ 2.16$&$ 2.55\pm 1.05$&$ 5.75\pm 0.76$&$ 144\pm 13$&$  51\pm  72$&$ 282\pm 31$&$0.33\pm 0.16$\\[0.8ex]
E56: GCIRS 34W        &$  1.5\pm 4.7$&$-4.86$&$ 1.91$&$ 1.86\pm 1.06$&$ 5.66\pm 0.43$&$ 157\pm 12$&$ -83\pm  79$&$ 323\pm 34$&$0.38\pm 0.16$\\[0.8ex]
E57                   &$  3.0\pm 6.9$&$ 5.30$&$ 0.30$&$ 1.78\pm 1.15$&$ 5.79\pm 0.62$&$  24\pm 11$&$ -33\pm  65$&$ 240\pm 35$&$0.45\pm 0.16$\\[0.8ex]
E61: GCIRS 34NW       &$ 10.3\pm 5.5$&$-4.48$&$ 3.42$&$ 3.70\pm 0.97$&$ 7.33\pm 1.04$&$ 132\pm 10$&$  -4\pm  63$&$ 283\pm 29$&$<0.46$\\[0.8ex]
\hline
\end{tabular}
}

~
\vspace*{0.2cm}

Positions are offsets from \Sgr. \\
$x$, $y$, $z$, $r$ are in units of $10^{17}$~cm, 
$v_r$ and $v_\phi$ --- in km/s, $\psi$ and $\phi$ --- in degrees.\\
In order to express distances in arcseconds one should
divide $x$, $y$, $z$, $r$ in the table by 1.20.\\
\end{table}

\end{document}